\documentclass[12pt]{article}
\usepackage[auth-sc]{authblk}
\usepackage{amsfonts,amssymb,amsmath,latexsym,logic,color,array}
\usepackage[mathscr]{eucal}
\usepackage{enumerate}
\usepackage{geometry}
\usepackage{graphicx}
\usepackage{epsf}

\newtheorem{theorem}{Theorem}[section]

\renewcommand{\next}{\!\raisebox{-.2ex}{ 
            \mbox{\unitlength=0.9ex
            \begin{picture}(2,2)
            \linethickness{0.06ex}
            \put(1,1){\circle{2}} 
            \end{picture}}}       
            \,}

\begin{document}

\title{On complexity of propositional Linear-time Temporal Logic with
  finitely many variables\thanks{Prifinal version of the paper
    published in: In van Niekerk J., Haskins
    B. (eds). \textit{Proceedings of SAICSIT'18}. ACM,
    2018. pp. 313-316. DOI: 10.1145/3278681.3278718}}
\author[1]{Mikhail Rybakov} \author[2]{Dmitry Shkatov} \affil[1]{Tver
  State University and University of the Witwatersrand, Johannesburg}
\affil[2]{University of the Witwatersrand, Johannesburg}
\maketitle

\begin{abstract}
  It is known~\cite{DS02} that both satisfiability and model-checking
  problems for propositional Linear-time Temporal Logic, {\bf LTL},
  with only a single propositional variable in the language are
  PSPACE-complete, which coincides with the complexity of these
  problems for {\bf LTL} with an arbitrary number of propositional
  variables~\cite{SislaClarke85}.  In the present paper, we show that
  the same result can be obtained by modifying the original proof of
  PSPACE-hardness for {\bf LTL} from~\cite{SislaClarke85}; i.e., we
  show how to modify the construction from~\cite{SislaClarke85} to
  model the computations of polynomially-space bound Turing machines
  using only formulas of one variable.  We believe that our
  alternative proof of the results from~\cite{DS02} gives additional
  insight into the semantic and computational properties of {\bf LTL}.
\end{abstract}
\maketitle

\section{Introduction}
\label{sec:intro}

The propositional Linear-time Temporal Logic {\bf LTL}, proposed
in~\cite{Pnueli77}, is historically the first temporal logic to have
been used in formal specification and verification of (parallel)
non-terminating computer programs~\cite{HR04}, such as (components of)
operating systems.  It has stood the test of time, despite a dizzying
variety of temporal logics that have since been introduced for the
purpose (see, e.g.,~\cite{DGL16}).

The task of verifying that a program conforms to a specification can
be carried out by checking whether an {\bf LTL} formula expressing the
specification is satisfied in the structure modelling the execution
paths of the program.  This corresponds to the model checking problem
for {\bf LTL}: given a formula, a model, and a state, check if the
formula is satisfied by all paths of the model beginning with the
given state.  The related task of verifying that a specification of a
program is consistent---and, thus, can be satisfied by some
program---corresponds to the satisfiability problem for {\bf LTL}:
given a formula, check whether there is a model and a path satisfying
the formula.

Therefore, the complexity of both satisfiability and model checking
are of crucial interest when it comes to applications of {\bf LTL} to
formal specification and verification.  It has been shown
in~\cite{SislaClarke85} that both satisfiability and model checking
for {\bf LTL} are PSPACE-complete.  It might have been hoped that the
complexity of satisfiability, as well as of model checking, may be
reduced if we consider a language with only a finite number of
propositional variables, which is sufficient for most
applications---as has been observed in~\cite{DS02}, most properties of
interest can be specified using a very small number of variables;
typically, not more than three.  Indeed, examples are known of logics
whose satisfiability problem goes down from ``intractable'' to
``tractable'' once we place a limit on the number of propositional
variables allowed in the language: thus, satisfiability for the
classical propositional logic and all the normal extensions of the
modal logic {\bf K5}~\cite{NTh75}, including logics such as {\bf K45},
{\bf KD45}, and {\bf S5} (see also~\cite{Halpern95}), used in formal
specification and verification of distributed and multi-agent
systems~\cite{FHMV95}, goes down from NP-complete to polynomial-time
decidable once we limit the number of propositional variables by an
(arbitrary) finite number.  Similarly, as follows
from~\cite{Nishimura60}, satisfiability for the intuitionistic
propositional logic goes down from PSPACE-complete to polynomial-time
if we allow only a single propositional variable in the language.

It has been shown in~\cite{DS02}, however, that even a single variable
in the language of {\bf LTL} is sufficient to produce a fragment whose
model-checking and satisfiability problems are as hard as
corresponding problems for the entire logic.  Thus, the complexity of
these tasks for {\bf LTL} cannot be lowered by placing restrictions on
the number of variables allowed in the construction of formulas.

It is often instructive to have various proofs of important formal
results, to which we believe the results on complexity of
model-checking and satisfiability for {\bf LTL} undoubtedly
belong. Thus, in the present paper, we present an alternative proof,
which is, in fact, a modification of the original proof
from~\cite{SislaClarke85} establishing PSPACE-hardness of {\bf LTL}
with an unlimited number of variables.  We show that, with some
ingenuity, one can modify the construction used
in~\cite{SislaClarke85} of an {\bf LTL}-model based on a computation
of a polynomially-space bound Turing machine so that we obtain a model
for a single-variable fragment of {\bf LTL}.  The interesting feature
of a modified construction is that---even though the size of the
model, and the {\bf LTL}-formula describing it, blows up---the blow-up
is proportionate to the size of the Turing machine, which is
independent of the size of the input, and thus the reduction remains
polynomial.

It is worth noticing that most well-know general methods
\cite{Halpern95, ChRyb03, RSh18a, RSh18b, RSh18c} of establishing
similar results for modal and temporal logics are not applicable to
{\bf LTL} due to the restiction on the branching factor in its models.

The paper is structured as follows. In
section~\ref{sec:syntax-semantics}, we briefly recall the syntax and
semantics of {\bf LTL}. In section~\ref{sec:single-variable-fragment},
we present our proof of PSPACE-hardness of model-checking and
satisfiability problems for the single-variable fragment of {\bf LTL},
which is a modification of the construction from the original proof
from~\cite{SislaClarke85}.  We conclude in
section~\ref{sec:conclusion} by drawing attention to some features of
{\bf LTL} that make it stand apart from other modal and temporal
logics used in formal specification and verification.

\section{Syntax and semantics}
\label{sec:syntax-semantics}

The language of {\bf LTL} contains an infinite set of propositional
variables $\Var = \linebreak \{p_1, p_2, \ldots \}$, the Boolean
constant $\bottom$ (``falsehood''), the Boolean connective $\imp$
\linebreak (``if \ldots, then \ldots''), and the temporal operators
$\next$ (``next'') and $\until$ (``until'').  The formulas are defined
by the following Backus-Naur form expression:
\[\vp ::= p  \mid\ \bottom\ \mid  (\vp
\imp \vp) \mid \next \vp \mid (\vp \until \vp), \]
where $p$ ranges over \Var.  We also define
$\top := ({\bottom} \imp {\bottom})$,
$\neg \vp := (\vp \imp {\bottom})$,
$(\vp \con \psi) := \neg (\vp \imp \neg \psi)$,
$\Diamond \vp := (\truth \until \vp)$, and
$\Box \vp := \neg \Diamond \neg \vp$.  We adopt the usual conventions
about omitting parentheses.  For every formula $\vp$ and every number
$n$ such that $n \geqslant 0$, we inductively define the formula
$\next^n \vp$ as follows: $\next^0 \vp := \vp$, and
$\next^{n+1} \vp := \next \next^n \vp$.

Formulas are evaluated in Kripke models (often referred to as
``transition systems'').  A Kripke model is a tuple
$\mmodel{M} = (\states{S}, \ar, V)$, where \states{S} is
a non-empty set \linebreak (of states), $\ar$ is a binary
(transition) relation on \states{S} that is serial (i.e., for every
$s \in \states{S}$, there exists $s' \in \states{S}$ such that
$s \ar s'$), and $V$ is a (valuation) function
$V: \Var \rightarrow 2^{\states{S}}$.

An infinite sequence $s_0, s_1, \ldots$ of states of \mmodel{M} such
that $s_i \ar s_{i+1}$, for every $i \geqslant 0$, is called a
\textit{path}.  Given a path $\pi$ and some $i \geqslant 0$, we denote
by $\pi[i]$ the $i$th element of $\pi$ and by $\pi[i, \infty]$ the
suffix of $\pi$ beginning with its $i$th element.

Formulas are evaluated with respect to paths. The satisfaction
relation between models $\frak{M}$, paths $\pi$, and formulas $\vp$ is
defined inductively, as follows:
\begin{itemize}
\item \sat{M}{\pi}{p_i} \sameas\ $\pi[0] \in V(p_i)$; \nopagebreak[3]
  \nopagebreak[3]
\nopagebreak[3]
\item \sat{M}{\pi}{\falsehood} never holds;
\nopagebreak[3]
\item \sat{M}{\pi}{(\vp_1 \imp \vp_2)} \sameas\ \sat{M}{\pi}{\vp_1}
  implies \sat{M}{\pi}{\vp_2}; \nopagebreak[3]
\item \sat{M}{\pi}{\next \vp_1} \sameas\ \sat{M}{\pi[1, \infty]}{\vp_1};
\nopagebreak[3]
\item \sat{M}{\pi}{\vp_1 \until \vp_2} \sameas\ \sat{M}{\pi[i,
    \infty]}{\vp_2}, for some $i \geqslant 0$, and \sat{M}{\pi[j,
    \infty]}{\vp_1} for every $j$ such that $0 \leqslant j < i$.
\end{itemize}
A formula is satisfiable if it is satisfied by some path of some
model. A formula is valid if it is satisfied by every path of every
model.

We now state the two computational problems considered in the
following section.  The {\it satisfiability problem for {\bf LTL}}:
given a formula $\vp$, determine whether there exists a model
$\frak{M}$ and a path $\pi$ in $\frak{M}$ such that \sat{M}{\pi}{\vp}.
The {\it model-checking problem for {\bf LTL}}: given a formula $\vp$,
a model $\frak{M}$, and a state $s$ in $\frak{M}$, determine whether
\sat{M}{\pi}{\vp} for every path $\pi$ such that $\pi[0] =
s$.
Clearly, formula $\vp$ is valid if, and only if, $\neg \vp$ is not
satisfiable; thus any deterministic algorithm that solves the
satisfiability problem also solves the validity problem, and vice
versa.



\section{Complexity of satisfiability and model-checking for
  finite-variable fragments}
\label{sec:single-variable-fragment}

In this section, we show how the original construction used in
~\cite{SislaClarke85} to establish PSPACE-hardness of model-checking
and satisfiability for {\bf LTL} with an arbitrary number of
propositional variables can be modified to prove that model-checking
and satisfiability for the single-variable fragments of {\bf LTL} are
PSPACE-hard, too. Before doing so, we briefly note that, for the
variable-free fragment, both problems are polynomially decidable.
Indeed, it is easy to check that every variable-free {\bf LTL} formula
is equivalent to either $\bottom$ or $\top$ (for example,
${\top} \until {\top}$ is equivalent to $\top$ and
${\top} \until {\bottom}$ is equivalent to $\bottom$); thus, to check
for satisfiability of a variable-free formula $\vp$, all we need to do
is to recursively replace each subformula of $\vp$ by either $\bottom$
or $\top$, which is linear in the size of $\vp$; likewise for
model-checking.




We recall that in~\cite{SislaClarke85} an arbitrary problem
``$x \in A?$'' solvable by polynomially-space bounded (deterministic)
Turing machines is reduced to model-checking for {\bf LTL}.  (The
authors of~\cite{SislaClarke85} then reduce the model-checking problem
for {\bf LTL} to the satisfiability problem for {\bf LTL}.) We show
how one can modify the construction from~\cite{SislaClarke85} to
simultaneously reduce the problem ``$x \in A?$'' to both model
checking and satisfiability for {\bf LTL} using formulas containing
only one variable.  Since we are describing a modification of a
well-known construction, we will be rather brief.  As we go along, we
point out the main differences of our construction from that
in~\cite{SislaClarke85}.

Let $M = (Q, \Sigma, q_0, q_1, a_0, a_1, \delta)$ be a (deterministic)
Turing machine, where $Q$ is the set of states, $\Sigma$ is the
alphabet, $q_0$ is the starting state, $q_1$ is the final state, $a_0$
is the blank symbol, $a_1$ is the symbol marking the leftmost cell,
and $\delta$ is the machine's program.  We adopt the convention that
$M$ gives a positive answer if, at the end of the computation, the
tape is blank save for $a_0$ written in the leftmost cell.  We assume,
for technical reasons, that $\delta$ contains an instruction to the
effect that the ``yes'' configuration yields itself (thus, we assume
that all computations with a positive answer are infinite). Given an
input on length $n$, we assume that the amount of space $M$ uses is
$S(n)$, for some polynomial $S$.

We now construct, in time polynomial in the size of $x$, a model
$\frak{M}$, a path $\pi$ in $\frak{M}$, and a formula $\psi$---of a
single variable, $p$---such that $x \in A$ if, and only if,
\sat{M}{\pi}{\vp}. It will also be the case that $x \in A$ if, and
only if, $\psi$ is {\bf LTL}-valid. The model $\frak{M}$ intuitively
corresponds, in the way described below, to the computation of $M$ on
input $x$.

First, we need the ability to model natural numbers within a certain
range, say $1$ through $k$.  To that end, we use models based on the
frame $\frak{F}_k$, depicted in Figure~\ref{fig:frame-k}, which is a
line made up of $k$ states.  By making $p$ true exactly at the $i$th
state of $\frak{F}_k$, where $1 \leqslant i \leqslant k$, we obtain a
model representing the natural number $i$.  We denote the model
representing the number $m$ by $\frak{N}_m$.

\begin{figure}
  \centering
  \begin{picture}(200,20)
  
\put(10,10){$\bullet$}
\put(40,10){$\bullet$}
\put(70,10){$\bullet$}
\put(160,10){$\bullet$}
\put(190,10){$\bullet$}

\put(14,12.5){\vector(1,0){27}}
\put(44,12.5){\vector(1,0){27}}
\put(74,12.5){\vector(1,0){27}}
\put(134,12.5){\vector(1,0){27}}
\put(164,12.5){\vector(1,0){27}}

\put(112.5,12){\ldots}
\end{picture}
\caption{Frame $\frak{F}_k$}
\label{fig:frame-k}
\end{figure}

We next use models $\frak{N}_m$ to build a model representing all
possible contents of a single cell of $M$.  Let $|Q| = n_1$ and
$|\Sigma| = n_2$.  As each cell of $M$ may contain either a symbol
from $\Sigma$ or a sequence $qa$, where $q \in Q$ and $a \in \Sigma$,
indicating that $M$ is scanning the present cell, where $a$ is
written, there are $n_2 \times (n_1 + 1)$ possibilities for the
contents of a single cell of $M$.  Let $k = n_2 \times (n_1 + 1)$;
clearly, $k$ is independent of the size of the input $x$.  To model
the contents of a single cell, we use models $\frak{N}_1$ through
$\frak{N}_k$ to build a model $\frak{C}$, depicted in
Figure~\ref{fig:model-c}, where small boxes represent models
$\frak{N}_1$ through $\frak{N}_k$.  In Figure~\ref{fig:model-c}, an
arrow from $s_0$ to a box corresponding to the model $\frak{N}_m$
represents a transition from $s_0$ to the first state of $\frak{N}_m$,
and an arrow from a box corresponding to the model $\frak{N}_m$ to
$s_1$ represents a transition from the last state of $\frak{N}_m$ to
$s_1$.  On the states in $\frak{N}_m$ ($1 \leqslant m \leqslant k$),
the evaluation of $p$ in $\frak{C}$ agrees with the evaluation of $p$
in $\frak{N}_m$; in addition, $p$ is false both at $s_0$ and $s_1$.

\begin{figure}
  \centering
  \begin{picture}(150,150)

    \put(40,70){$\bullet$}
    \put(30,70){$s_0$}
    \put(103,70){$\bullet$}
    \put(110,70){$s_1$}
\put(70,12){$\Box$}
\put(70,97){$\Box$}
\put(70,125){$\Box$}

\put(42.5,72.5){\vector(1,1){28}}
\put(42.5,72.5){\vector(1,2){28}}
\put(42.5,72.5){\vector(1,-2){28}}

\put(76.5,128){\vector(1,-2){27}}
\put(76.5,99){\vector(1,-1){27}}
\put(76.5,16){\vector(1,2){27}}

\put(79.5,128){$\frak{N}_1$}
\put(79.5,99){$\frak{N}_2$}
\put(69,59){$\ldots$}
\put(79.5,11){$\frak{N}_{k}$}
\end{picture}
  \caption{Model $\frak{C}$.}
  \label{fig:model-c}
\end{figure}

Let the length of $x$ be $n$.  We use $S(n)$ copies of $\frak{C}$ to
represent a single configuration of $M$.  This is done with the model
$\frak{M}$, depicted in Figure~\ref{fig:model-T}.  In $\frak{M}$, a
chain made up of $S(n)$ copies of $\frak{C}$ is preceded by a model
$\frak{B}$ marking the beginning of a configuration; the use of
$\frak{B}$ allows us to separate configurations from each other. All
that is required of the shape of $\frak{B}$ is for it to contain a
pattern of states (with an evaluation) that does not occur elsewhere
in $\frak{M}$; thus, we may use the frame $\frak{F}_3$ and define the
evaluation to make $p$ true at its every state.

This completes the construction of the model $\frak{M}$.  One might
think of $\frak{M}$ as consisting of ``cycles,'' each cycle
representing a single configuration of $M$ in the following way: to
obtain a particular configuration of $M$, pick a path from the first
state of $\frak{B}$ to the last state of the last copy of $\frak{C}$
that traverses the model $\frak{N}_i$ withing the $j$th copy of
$\frak{C}$ exactly when the $j$th cell of the tape of $M$ contains the
$i$th ``symbol'' from the alphabet $\Sigma \union Q \times \Sigma$.

The main, and crucial, difference between the model $\frak{M}$
described above and the model used in~\cite{SislaClarke85} is that we
use ``components'' $\frak{N}_k$ where~\cite{SislaClarke85} use an
anti-chain of $k$ states distinguished by the evaluation of $k$
distinct propositional variables.  This allows us---in contrast
to~\cite{SislaClarke85}---to use a single propositional variable in
describing our model.

\begin{figure}
  \centering
\begin{picture}(200,50)
\put(10,19){$\Box$}
\put(40,19){$\Box$}
\put(70,19){$\Box$}
\put(160,19){$\Box$}
\put(190,19){$\Box$}

\put(16.5,22.5){\vector(1,0){24.5}}
\put(46.5,22.5){\vector(1,0){24.5}}
\put(76.5,22.5){\vector(1,0){24.5}}
\put(136.5,22.5){\vector(1,0){24.5}}
\put(166.5,22.5){\vector(1,0){24.5}}

\put(112.5,22){\ldots}

\put(193.5,20){\vector(0,-1){10}}
\put(193.5,10){\vector(-1,0){180}}
\put(13.5,10){\vector(0,1){10}}

\put(10,29){$\frak{B}$}
\put(40,29){$\frak{C}$}
\put(70,29){$\frak{C}$}
\put(160,29){$\frak{C}$}
\put(190,29){$\frak{C}$}

\end{picture}
\caption{Model $\frak{M}$}
\label{fig:model-T}
\end{figure}

We now describe how to build a formula $\psi$ whose satisfaction we
want to check with respect to an infinite path beginning with the
first state of $\frak{B}$.  It is rather straightforward to write out
the following formulas (all one needs to say is what symbols are
written in each of the cells of $M$'s tape):
\begin{itemize}
\item A formula $\psi_{start}$ describing the initial configuration
  of $M$ on $x$;
\item A formula $\psi_{positive}$ describing the configuration of $M$
  corresponding to the positive answer.
\end{itemize}
The length of both $\psi_{start}$ and $\psi_{positive}$ is clearly
proportionate to $k \times S(n)$, as we have $S(n)$ cells to describe
and use formulas of length proportionate to $k$ to describe each of
them.  Next, we can write out a formula $\psi_{\delta}$ describing the
program $\delta$ of $M$.  This can be done by starting with formulas
of the form $\next^j \sigma$, where $j$ is the number of states in a
path leading from the first state of $\frak{B}$ to the last state of
the last copy of $\frak{C}$ in a single ``cycle'' in $\frak{M}$, to
describe the change in the contents of the cells from one
configuration to the next, and then, for each instruction $I$ from
$\delta$, writing a formula $\alpha(I)$ of the form
$\bigwedge_{i=0}^{S(n)} \Box \chi$, where $\chi$ describes changes
occurring in each cell of $M$.  Clearly, the length of each
$\alpha(I)$ is proportionate to $k \times S(n)$.  Then,
$\psi_{\delta} = \bigwedge_{I \in \delta} \alpha(I)$.  As the number
of instructions in $\delta$ is independent from the length of the
input, the length of $\psi_{\delta}$ is proportionate to
$c \times S(n)$, for some constant $c$.

Lastly, we define
$$
\psi = \psi_{start} \con \Box \psi_{\delta} \imp \Diamond
\psi_{positive}.
$$
One can then show, by induction on the length of the computation of
$M$ on $x$, that $M(x) = yes$ if, and only if, $\psi$ is satisfied in
$\frak{M}$ by an infinite path corresponding, in the way described
above, to the computation of $M$ on $x$.  This gives us the following:
\begin{theorem}
  The model-checking problem for {\bf LTL} formulas with at most one
  variable is \rm{PSPACE}-complete.
\end{theorem}
Likewise, we can show that $M(x) = yes$ if, and only if, $\psi$ is
satisfiable, which gives us the following:
\begin{theorem}
  The satisfiability problem for {\bf LTL} formulas with at most one
  variable is \rm{PSPACE}-complete.
\end{theorem}

\section{Conclusion}
\label{sec:conclusion}

We have shown how the construction from~\cite{SislaClarke85} can be
modified to prove the PSPACE-hardness of both model-checking and
satisfiability for the single-variable fragment of the propositional
Linear-time Temporal Logic, {\bf LTL}.  The essential difference
between the original construction and the modified construction
presented above is that we use chains of states of length
$n_2 \times (n_1 + 1)$, where $n_1$ and $n_2$ are the number of states
and symbols, respectively, of the Turing machine whose computation we
model, rather than single states used in~\cite{SislaClarke85} to
evaluate $n_2 \times (n_1 + 1)$ variables.  Since numbers $n_1$ and
$n_2$ are not known in advance, the modelling in~\cite{SislaClarke85}
requires an unlimited number of variables.  In our modification of the
proof from~\cite{SislaClarke85}, the number $n_2 \times (n_1 + 1)$ is
reflected in the model that can be described by formulas with a single
variable, thus producing a reduction to a single-variable formula.
Even though the length of the formula is clearly dependent on
$n_2 \times (n_1 + 1)$, this number is independent of the input $x$ to
the problem ``$x \in A?$'' which we are reducing to the model-checking
and satisfiability for {\bf LTL}; thus, the reduction remains
polynomial.  This is a rather curious property of {\bf LTL}, which
makes it stand apart from most ``natural'' modal and temporal logics
(by a ``natural'' logic, we mean a logic that was not purposefully
constructed to exhibit a certain property).

We conclude by drawing attention to another peculiarity of {\bf LTL}
that makes it stand apart from other ``natural'' modal and temporal
logics.  While the complexity function (see~\cite{ChZ97}, Section
18.1) for {\bf LTL}, both in the language with infinitely many
variables, and---as follows from the proof presented above, in the
language with a single variable---is polynomial, the complexity of the
corresponding satisfiability problem is PSPACE-complete. By contrast,
for most ``natural'' modal and temporal logics, the polynomiality of
the complexity function implies the polynomial-time decidable
satisfiability problem, and PSPACE-completeness of satisfiability
problem implies the exponential complexity function.

\end{document}